\documentstyle[12pt,epsfig]{article}

\def\gsim{\lower0.5ex\hbox{$\:\buildrel >\over\sim\:$}}
\def\lsim{\lower0.5ex\hbox{$\:\buildrel <\over\sim\:$}}

\def \sneub{{\bar {\tilde\nu}\hspace{-1mm}\stackrel{\mu}{\rule{0mm}{0.5mm}}}}
\def \sneu{{\tilde\nu}^\mu}
\def \sneut{{\tilde\nu}^\tau}
\def \sneue{{\tilde\nu}^e}

\def \sneui{{\tilde\nu}^i}

\def \sneup{{\tilde\nu}_+}
\def \sneum{{\tilde\nu}_-}
\def \sneupm{{\tilde\nu}_{\pm}}
\def \rp{{R\hspace{-0.22cm}/}_P}
\def \lp{{L\!\!\!/}}

\def \be {\begin{equation}}
\def \ee {\end{equation}}
\def \bea{\begin{eqnarray}}
\def \eea{\end{eqnarray}}

\def \n{\noindent}

\begin{document}
\baselineskip 18pt plus 2pt

\noindent \hspace*{10cm}UCRHEP-T221 (April 1998)\\
\noindent \hspace*{10cm}BNL-HET-98/16 

\begin{center}
{\bf \boldmath ${R}$--parity violation, sneutrino mixing phenomena and 
CP-violation in \boldmath ${p \bar p \to \tilde\nu \to \ell^+ \ell^- + X}$}
\vspace{.2in}

S. Bar-Shalom, G. Eilam\footnote{On leave from: Physics Department, 
Technion-Institute of Technology, Haifa 32000, 
Israel.}\\
\vspace{-5pt} 
Physics Department, University of California, Riverside, 
CA 92521, USA\\
\bigskip

and A. Soni\\
\vspace{-5pt}
Physics Department, Brookhaven National Laboratory, Upton, NY 11973,
USA\\
\vspace{.2in}

{\bf Abstract}\\
\end{center}

\n The sneutrino resonance reactions 
$p \bar p \to \tilde\nu +X \to \ell^+ \ell^- +X$, for $\ell=e,\mu,\tau$,
 in the MSSM without 
$R$-parity are considered. We perform a cross section analysis and 
show that present limits on some products of $\rp$ couplings 
in the sneutrino sector can be significantly improved in  future 
upgraded Tevatron runs. Furthermore, we introduce CP-violating 
and CP-conserving $\tau$-spin asymmetries 
which are generated already at the {\it tree-level} in the reaction 
$p \bar p \to \tilde\nu +X \to \tau^+ \tau^- +X$ if 
there is $\sneu$-$\sneub$ mixing and that vanish  
in the SM\null. We find, for example, that for muon-sneutrino 
masses in the range 
$150~{\rm GeV}\lsim m_{{\tilde\nu}^\mu} \lsim 450~{\rm GeV}$,     
these spin asymmetries  reach $\sim 20$--30\% and $\sim 10\%$ 
for mass splitting between the muon-sneutrino CP-odd and CP-even 
states at the level of 
$\Delta m \sim \Gamma$ and $\Delta m \sim \Gamma/4$, respectively,
where $\Gamma$ is the ${\tilde\nu}^\mu$ width.
Both the CP-violating and the CP-conserving 
spin asymmetries should be detectable in future Tevatron runs  
even for a heavy sneutrino with a mass $\lsim 500$ GeV\null.
If detected, such asymmetries---being proportional to the mass splitting 
between the CP-even and CP-odd sneutrino states---may serve as a
strong indication for the existence of the sneutrino mixing phenomenon. 
\newpage

\n \underline{\bf 1. Introduction and working assumptions}\\

\n The minimal supersymmetric standard model (MSSM)
 with an $R$-parity conserving superpotential 
possesses a very distinct prediction:
superpartners must be produced in pairs and, as a 
consequence, the lightest supersymmetric particle is stable. 
This implies that direct production of sparticles is 
restricted to colliders with a c.m.\ energy at least twice 
the typical sparticle mass. 
Although the imposition of $R$-parity conservation on the MSSM Lagrangian 
was originally proposed in order to avoid fast proton decays 
\cite{rproton}, it is well known that if the $\rp$ Lagrangian 
violates only lepton number (or only baryon number), then      
the proton lifetime does not pose any phenomenological problem.
Since in supersymmetry (SUSY) models, 
the supermultiplets of the lepton-doublet 
${\hat L}$ and the down-Higgs doublet ${\hat H}_d$ have the same 
gauge quantum numbers, 
the $\rp$ lepton number violating operators are constructed simply  
by the replacement ${\hat H}_d \to {\hat L}$ \cite{rpreview,reffoot1}:

\be
W_{\lp} = \frac{1}{2}\lambda_{ijk} {\hat L}_i  
{\hat L}_j {\hat E}_k^c + \lambda_{ijk}^{\prime} {\hat L}_i  
{\hat Q}_j {\hat D}_k^c
\label{rparity}~,     
\ee

\n where ${\hat L}$ and ${\hat Q}$ are the SU(2)-doublet lepton and quark 
superfields and ${\hat E}^c$ and ${\hat D}^c$ are the 
lepton and quark singlet 
superfield. Also, $i,j,k$ are flavor indices such that, 
in the pure leptonic operator in Eq.~\ref{rparity},  
$i \neq j$. The presence of $W_{\lp}$ drastically changes the 
phenomenology of the SUSY 
leptonic sector since it gives rise to the possibility of 
having $s$-channel slepton resonant formation in scattering 
processes, thus enabling the detection of sleptons with masses roughly 
up to the collider c.m.\
energy \cite{resold,resnew,zerwaspp,hepph9802251}. 

In this work we focus specifically on the effects of the $\rp$ interactions 
in Eq.~\ref{rparity} on lepton-pair production processes at the Fermilab 
Tevatron, 
$p \bar p \to \ell^+ \ell^- +X$, with $\ell=e,\mu,\tau$. 
In particular, in the presence of these new $\rp$ couplings there is an 
additional (apart from the SM) contribution to the total cross section 
coming from $\hat s$-channel sneutrino resonances \cite{zerwaspp,foot1}. 
We will show below how the future Tevatron run with c.m.\ energy 
$\sqrt{s}=2$ TeV and an integrated 
luminosity of ${\cal L}=2$ fb$^{-1}$ \cite{tevatron} 
is capable of significantly improving 
the existing limits on some products, $\lambda \lambda^{\prime}$, 
of $\rp$ couplings through a detailed study of the 
total cross sections for $p \bar p \to \ell^+ \ell^- +X$.     

Apart from the possibility of constraining the $\rp$ couplings 
which enter the sneutrino resonant formation in 
$\ell$-pair production, we will also explore two new 
aspects of $\sneu$ resonance at the Tevatron: the detection of 
$\sneu - \sneub$ mixing and CP-violation in the process 
$p \bar p \to \tau^+ \tau^- +X$. Both  
phenomena may exist once $\lambda_{323},\lambda_{ijk}^{\prime} \neq 0$ 
in Eq.~\ref{rparity}. In a previous work \cite{hepph9802251}, 
we have presented a detailed investigation of these two new issues for 
the process $\ell^+ \ell^- \to \tau^+ \tau^-$. We showed there 
that the sneutrino mixing 
phenomenon as well as the CP-violating effects, may be  easily detected 
already at the CERN $e^+e^-$ collider LEP2 if indeed the pure leptonic 
sneutrinos $\rp$ couplings in Eq.~\ref{rparity}
 exist.       
In what follows, we will present an analogous investigation appropriate 
for the reaction $p \bar p \to \tau^+ \tau^- +X$ at the Tevatron. 
In leptonic colliders, 
where the c.m.\ energy is fixed by the energy of the colliding leptons, 
in order to discover a new resonance particle 
one is forced to tune the c.m.\ energy to the new particle mass 
which is of course not possible if the particle mass is a priori unknown.  
The advantage of a hadron collider is that,   
due to the continuous energy distribution of the colliding partons, 
one can probe new resonances over a much wider 
range of the corresponding new particle mass.    
We will therefore show that, contrary to LEP2 where the new effects 
mentioned above can be detected only if the sneutrino mass lies within a 
$\sim 10$ GeV range of the LEP2 c.m.\ energy, at the Tevatron 
these effects can be detected over a $\sim 300$ GeV sneutrino 
mass range if $\tau$-polarization could be measured. 

Sneutrino mixing phenomena have been gaining some 
interest recently \cite{mix1,mixyuval,mix2}.
The question of whether sneutrinos mix or not is of 
fundamental importance since this mixing is closely 
related to the generation of neutrino masses \cite{mix1,mixyuval}. 
Here we are interested in the detection of sneutrino 
mixing rather than in its origins. We therefore do not 
assume any specific model for it to occur. Instead, for a given 
sneutrino flavor $i=e,\mu,\tau$ 
we write ${\tilde\nu}^i = (\sneup^i +i \sneum^i)/\sqrt 2$ and 
simply assume that, due to some new short distance physics, 
there is a mass splitting between 
the new CP-even and CP-odd sneutrino mass eigenstates 
$\sneup^i$ and $\sneum^i$, respectively 
(we assume CP conservation in the mixing). 
In fact, it was found in \cite{mixyuval} that, generically, 
$\Delta m_{\sneupm^i} / m_{\nu_i} \lsim {\rm few} 
\times 10^3$ is roughly required in order for the 
neutrino masses, $m_{\nu_i}$,
 to be within their present experimental upper limits. 
One therefore expects the mass splitting in the $\sneupm^e$ sector to 
be extremely small. But, for $\sneupm^\mu$ (and in particular for 
$\sneupm^\tau$, which is however not of our main interest here,
insofar as the issue of mixing goes) the mass splitting 
can be sizable enough to drive significantly large 
 new CP-odd and CP-even asymmetries already at the tree-level in 
$\tau$-pair production at leptonic colliders \cite{hepph9802251} 
and at the Tevatron.     

Let us now establish our working assumptions and conventions: 
here and throughout the rest of the paper we assume for simplicity and 
without loss of generality that the $\rp$ couplings  
$\lambda^{\prime}_{ijk}$ and $\lambda_{pip}$ for all 
allowed combinations of indices,  except for $\lambda_{323}$, are  real   
(this assumption does not affect the calculations presented 
in this paper). As will be shown in section 3,  
the imaginary part of $\lambda_{323}$ can be responsible for 
large tree-level CP-violating effects in 
the reaction $p \bar p \to \tau^+ \tau^- +X$. 

Although the existence of sneutrino mixing is irrelevant for 
the purpose of the cross section analysis performed in the next section, 
for definiteness, total cross sections will  be calculated 
 in the $\sneupm^i$ mass basis. The relevant couplings 
of the CP-even ($\sneup^i$) and the CP-odd ($\sneum^i$) sneutrino 
mass eigenstates are then: 

\bea
d_j \sneup^i d_k = i \lambda^{\prime}_{ijk}/\sqrt 2 ~~,~~ 
d_j \sneum^i d_k = - \lambda^{\prime}_{ijk} \gamma_5/\sqrt 2~,
\eea

\n and for  
$\left\{ i,p \right\} \neq \left\{ 2,3 \right\}$:

\bea 
\ell_p \sneup^i \ell_p = i \lambda_{pip}/\sqrt 2~~,~~ 
\ell_p \sneum^i \ell_p = - \lambda_{pip} \gamma_5/\sqrt 2~.
\eea
 
\n For $i=2,~p=3$ we define $\lambda_{323} \equiv (a+ib)/\sqrt 2$, 
therefore:  

\bea
\tau \sneup^\mu \tau = i (a-ib \gamma_5)/2~~,~~ 
\tau \sneum^\mu \tau = i (b+ia \gamma_5)/2~. 
\eea

As mentioned before, the CP-violating and CP-conserving asymmetries in the
$\tau$-pair production channel are proportional to  the possible 
mass splitting between $\sneup^\mu$ and $\sneum^\mu$ \cite{foot2}.
In section 3 we will specifically take
$\Delta m_{\sneupm^\mu} \lsim \Gamma_{\sneupm^\mu}$,  
where throughout the paper  
we set $\Gamma_{\sneupm^i}=10^{-2} m_{\sneupm^i}$ for 
any sneutrino flavor $i=e,\mu$ or $\tau$.   
Indeed, if  $m_{\sneupm^i} > m_{\tilde {\chi}^+},~m_{\tilde {\chi}^0}$ 
($\tilde{\chi}^+$ and $\tilde{\chi}^0$ are the chargino and neutralino, 
respectively), then the two-body decays 
$\sneupm^i \to \tilde{\chi}^+ \ell, ~\tilde{\chi}^0 \nu$ are open and the 
corresponding partial widths are given by 
(see Barger {\it et al.} in \cite{resold}):  

\bea
&&\Gamma(\sneupm^i \to \tilde{\chi}^+ \ell) 
\sim {\cal O} \left[ 10^{-2} m_{\sneupm^i} \times 
\left(1 - m_{\tilde {\chi}^+}^2/m_{\sneupm^i}^2 \right)^2 \right]~,
\label{eq5} \\
&&\Gamma(\sneupm^i \to \tilde{\chi}^0 \nu) 
\sim {\cal O} \left[ 10^{-2} m_{\sneupm^i} \times 
\left(1-m_{\tilde {\chi}^0}^2/m_{\sneupm^i}^2\right)^2 \right]~.
\label{eq6}
\eea 

\n In Eqs.~\ref{eq5} and \ref{eq6} above we have omitted factors
coming from the  charginos and neutralinos mixing matrices, respectively,
which  should multiply the right hand side of Eqs.~\ref{eq5} and \ref{eq6}.  
Since these mixing factors depend on the MSSM parameters we assume 
for simplicity that at least for one chargino and one neutralino they are 
of order unity. This assumption does not change our predictions in section 2.
In fact, if the mixing factors are much smaller than 1, then 
with a sneutrino mass splitting of the order of a GeV, our spin asymmetries 
in section 3 are even more useful for the detection of 
such a mass splitting. Assuming that at least one chargino and one neutralino 
has a mass below 100 GeV and summing the two partial widths, then
for $m_{\sneupm^i} \gsim 150$ GeV,  
$\Gamma_{\sneupm^i}= 10^{-2} m_{\sneupm^i}$ serves our purpose as it 
is a viable estimate even without taking into account the new $\rp$ 
two-body decay modes which, when summed, can form a significant 
fraction of the total sneutrino width. 

Our motivation for choosing the specific condition 
$\Delta m_{\sneupm^\mu} \lsim \Gamma_{\sneupm^\mu}$ 
for the $\sneupm^\mu$ mass splitting is twofold:

\begin{enumerate}
\item In such a case the two $\sneup^\mu$ and $\sneum^\mu$ resonances 
will overlap and  distinguishing between them becomes a 
non-trivial experimental task. Thus, the CP-conserving and CP-violating 
$\tau$-spin asymmetries presented in section 3 
may provide a feasible alternative for establishing that 
 $m_{\sneup^\mu} \neq  m_{\sneum^\mu}$. 
\item With $\Gamma_{\sneupm^i}= 10^{-2} m_{\sneupm^i}$, 
the choice $\Delta m_{\sneupm^\mu} \lsim \Gamma_{\sneupm^\mu}$ 
implies $\Delta m_{\sneupm^\mu} / m_{\sneupm^\mu} << 1$, as imposed  
by neutrino masses \cite{mixyuval,footbound}.
\end{enumerate}

We will show in section 3 that CP-odd and CP-even $\tau$-polarization
asymmetries at the level of tens of a percent 
may arise in the reaction $p \bar p \to \tau^+ \tau^- +X$ 
within the muon-sneutrino mass range 
$150 ~{\rm GeV} \lsim m_{\sneupm^\mu} \lsim 450 ~{\rm GeV}$, even 
for a mass splitting as small as 
$\Delta m_{\sneupm^\mu} = \Gamma_{\sneupm^\mu}/4$.     
These asymmetries (for $\Delta m_{\sneupm^\mu} = \Gamma_{\sneupm^\mu}/4$) 
are detectable at future Tevatron runs,  with a  statistical significance
above $3 \sigma$ throughout almost the  entire mass range 
$150 ~{\rm GeV} \lsim m_{\sneupm^\mu} \lsim 450$ GeV\null. 
If $\Delta m_{\sneupm^\mu} \sim \Gamma_{\sneupm^\mu}$ the effects are much more
pronounced.

It is especially interesting that a large tree-level 
CP-violating effect may emanate in the reaction 
$p \bar p \to \tau^+ \tau^- +X$ and, in particular, may be detectable at  
the future runs of the Tevatron.
Previous studies of CP-violating effects in 
$\tau^+ \tau^-$ final state, attributed to models beyond the SM  
such as multi-Higgs doublet model, SUSY, 
leptoquarks and Majorana $\nu$, all involve one-loop exchanges of 
the new particles which generate a CP-violating electric dipole moment 
for the $\tau$ (see \cite{cptau} and references therein). These CP-odd
effects are therefore much smaller than our tree-level effect. 

The paper is organized as follows: in section 2 we discuss the 
limits on some of the $\rp$ couplings  attainable at future runs of the 
Tevatron. In section 3 we calculate our CP-violating and CP-conserving 
$\tau$-polarization asymmetries and discuss the 
numerical results and in section 4 we summarize.\\
 
\n \underline{\bf 2. Expected new 
limits on \boldmath ${\rp}$ couplings at the future upgraded Tevatron}\\
  
\n In this section we perform a detailed investigation of the reaction 
$p \bar p \to \ell^+ \ell^- +X$ at the Tevatron. In the presence of 
the $\rp$ couplings in Eq.~\ref{rparity}, the cross section 
receives contributions from both the SM $\hat s$-channel 
$\gamma,Z$ exchanges and from the new $\hat s$-channel sneutrino 
exchanges \cite{foot1}.
The interferences between the SM diagrams 
and the $\hat s$-channel $\sneupm^i$ 
diagrams as well as between the $\sneup^i$ and 
the $\sneum^i$ $\hat s$-channel exchange diagrams are proportional to 
the down-quarks masses  
and are therefore being neglected.    
As a consequence, the total hard cross section, ${\hat \sigma}^T$, 
can be expressed as the incoherent sum of the SM and $\sneupm^i$ parton-level 
cross sections: ${\hat \sigma}^T = {\hat \sigma}_{SM}^{jk} +   
{\hat \sigma}_{\sneupm^i}^{jk}$. For the SM,   
${\hat \sigma}_{SM}^{jk} \equiv 
\sigma(q_j {\bar q}_k \to \gamma, ~Z \to \ell^+ \ell^-)$ such that  
$q=u~{\rm or}~d$, {\it i.e.}, up- or down-quark and $j=k$, 
where $j,k =1,2,3$ are flavor indices.   
For the $\hat s$-channel sneutrino case
 only down-quark annihilation contributes, since 
sneutrino couplings to up-quarks are forbidden by gauge invariance.
We, therefore, define:  
${\hat \sigma}_{\sneupm^i}^{jk} \equiv 
\sigma (d_j {\bar d}_k \to \sneupm^i \to \ell^+_p \ell^-_p)$, 
where we consider only flavor-diagonal lepton pair production and 
we explicitly keep the flavor indices $j,k,p=1,2,3$ as, in principle, 
all combinations of the $d_j \sneupm^i d_k$ and 
$\ell_p \sneupm^i \ell_p$ couplings are present in 
the lagrangian of Eq.~\ref{rparity}, with $p \neq i$.   

The total cross section $\sigma^T = \sigma(p \bar p \to \ell^+ \ell^- +X)$ 
can also be subdivided into the SM and the $\sneupm^i$ parts as  
$\sigma^T = \sigma^T_{SM} + \sigma^T_{\sneupm^i}$. Each part    
is then given in terms of the parton luminosities 
$d {\cal L}_{jk}(\tau)/
d \tau$ \cite{collider}:

\bea
\sigma^T_{SM};~\sigma^T_{\sneupm^i} = \sum_{j,k} \int_{\tau_-}^{\tau_+} 
d \tau ~ \frac{d {\cal L}_{jk}(\tau)}
{d \tau}~ {\hat \sigma}^{jk}_{SM};~{\hat \sigma}^{jk}_{\sneupm^i }
 (\hat s =\tau s) \label{sigpp}~,
\eea

\n where $\sqrt{\hat s}$ and $\sqrt{s}$ are the c.m.\ 
energies of the $q \bar q$ and 
$p \bar p$ systems, respectively. 
For later use, the integration over the variable $\tau$ is 
carried out by imposing 
 lower and upper cuts $\tau_-$ and $\tau_+$, respectively, corresponding 
to lower ($M_{\ell \ell}^-$) and upper ($M_{\ell \ell}^+$) 
cuts on the $\ell^+ \ell^-$ invariant mass 
$M_{\ell \ell} \equiv \sqrt {\hat s}$. Also: 

\bea
\frac{d {\cal L}_{jk}(\tau)}
{d \tau} = \frac{{\cal C}_{jk}}{1+\delta_{jk}} 
\int_\tau^1 \frac{d x_j}{x_j} 
\left[ f_{j/{\rm p}}(x_j) f_{k/\bar{\rm p}}(\tau/x_j) + 
(j \leftrightarrow k) \right] 
\label{partonlum}~,
\eea 
      
\n where the color factor ${\cal C}_{jk}$ 
arises from summing and averaging over initial colors and for our processes 
${\cal C}_{jk}=1/3$. For the distribution functions, $f_{j/{\rm p}}(x_j)$, 
we use the CTEQ4M parametrization \cite{cteq}.   

The SM parton-level cross section is given by 
(neglecting the quarks and leptons masses):

\bea
{\hat \sigma}_{SM}^{jk} &=& \frac{\pi \alpha^2}{3 \hat s} 
\left( s_W^4 c_W^4 \beta_Z \right)^{-1} \times 
\left\{4 Q_q^2 s_W^4 c_W^4 \beta_Z  - \right. \nonumber \\
&&\left. 2 Q_q s_W^2 c_W^2 (1-x_Z) (g_L+g_R) (g_L^q +g_R^q) + 
(g_L^2+g_R^2) ((g_L^q)^2+(g_R^q)^2) \right\} \label{sigsm}~,
\eea

\n where  
$\beta_Z \equiv (1-x_Z)^2 + x_Z^2 (\Gamma_Z/m_Z)^2$ and   
$x_Z\equiv m_Z^2/\hat s$. Also, 
$g_L=s_W^2 -1/2$, $g_R=s_W^2$ and  
$g_{L,R}^q=T_{L,R}^{3} - s_W^2 Q_q$. Here  
$s_W$ is the sine of the weak mixing angle $\theta_W$, $Q_q$ is the charge of 
$q$ and  
$T_{L,R}^{3}$ are the appropriate $z$-components of the weak isospin of 
the quark $q$.   

For the $\hat s$-channel sneutrino contribution
it is useful to write the hard cross section in the form:

\bea
{\hat \sigma}_{\sneui}^{jk} = v_{ip}^{jk} 
\times g^i(\hat s) \label{sigsneu0}~,
\eea
\n where:
\bea
\sqrt {v_{ip}^{jk}} \equiv |\lambda_{pip}| 
\lambda^{\prime}_{ijk} ~~,~~ g^i(\hat s) = 
\frac{\hat s}{32 \pi 
\left|\hat s -m_{\sneui}^2 +i m_{\sneui} 
\Gamma_{\sneui} \right|^{2}}~.
\eea 

\n In this section we assume $m_{\sneui} \equiv m_{\sneup^i}=m_{\sneum^i}$
and $\Gamma_{\sneui}  \equiv \Gamma_{\sneup^i}=\Gamma_{\sneum^i}$.
Therefore, through the present section starting with Eq.~\ref{sigsneu0}~,
we drop the subscript $\pm$, then resurrect it in the next section.
As mentioned above, we set 
$ \Gamma_{\sneui}=10^{-2}m_{\sneui}$.
It is then clear from Eqs.~\ref{sigpp}, \ref{partonlum} and 
\ref{sigsneu0} that  
the $\sneui$ contribution to the total cross section,  
{\it i.e.}, $\sigma^T_{\sneui}$, can be expressed as a sum over the 
$\rp$ couplings $v_{ip}^{jk}$ with corresponding weights $c_{jk}$ as:

\bea
\sigma^T_{\sneui} = \sum_{jk} c_{jk} v_{ip}^{jk} ~,
\eea

\n where:

\bea
c_{jk} \equiv \int_{\tau_-}^{\tau_+} d \tau ~ \frac{d {\cal L}_{jk}(\tau)}
{d \tau} g^i(\hat s) \label{cjk}~.
\eea 

\n In what follows we have employed an upper cut on 
the $\ell^+ \ell^-$ system 
invariant mass of $M_{\ell \ell}^+ =500$ GeV.
Also, the $c_{jk}$'s are calculated with  
the three lower cuts $M_{\ell \ell}^- =50,~100$ and 150 GeV.
The third lower cut, {\it i.e.}, $M_{\ell \ell}^- =150$ GeV,  
is useful for
removing most of the SM contribution via the $\hat s$-channel $Z$ 
exchange. We will show below that such a cut can substantially improve 
the limits on the $\rp$ couplings at the future 
Tevatron Run II 
through an analysis of the reaction $p \bar p \to \ell^+ \ell^- +X$.

In Tables 1, 2 and 3 we list the presently allowed upper limits on the 
products $\lambda \lambda^{\prime}$ relevant for $e,\mu$ and $\tau$-
pair production processes, respectively.   
We combine the limits on individual $\rp$ couplings 
\cite{rpreview} with existing limits on products of $\lambda \lambda^{\prime}$ 
\cite{foottables,productlimits}, where in Tables 1--3 we write the more 
stringent ones, {\it i.e.}, coming either from the individual or the 
$\lambda \lambda^{\prime}$ bounds \cite{reffoot2}.
Evidently, from the numbers in 
Tables 1--3, two useful conclusions can be drawn:

\begin{enumerate}
\item As expected, the limits on flavor changing couplings, 
{\it i.e.}, $j\neq k$, relevant  
for $e$ and $\mu$-pair production are much tighter, thus their 
contribution to the corresponding cross sections is negligible.
\item In the $\mu$ and $\tau$-pair production case the  
$\rp$ couplings associated with $\sneue$ 
are negligible compared to the other sneutrino flavor, {\it i.e.},  
$\sneut$ for the $\mu \mu$ case and $\sneu$ for 
$\tau \tau$ production. 
\end{enumerate}

\n These two distinct features come in extremely handy later on when we 
discuss the attainable limits on the products
$\lambda \lambda^{\prime}$ at the future Tevatron Run II\null. 
\begin{table}[htb]
\begin{center}
\begin{tabular}{||r||r|r||}  \hline \hline
\multicolumn{1}{||c||}{~~} & \multicolumn{2}{|c||}{~~} \\
\multicolumn{1}{||c||}{~~} & \multicolumn{2}{c||}{\raisebox{3mm}{$d_j 
{\bar d}_k \to \sneui \to e^+ e^-$}}\\ \cline{2-3}
$jk \Downarrow$ & $(\sneu)~~\lambda_{121} 
\lambda^{\prime}_{2jk}/n_\mu^2$ & $(\sneut)~~\lambda_{131} 
\lambda^{\prime}_{3jk}/n_\tau^2$ \\ \hline \hline
$11$& $4.5\times 10^{-3}$ & $6.0\times 10^{-3}$ \\ \hline 
$12,~21$& $2.5\times 10^{-8}$  & $2.5\times 10^{-8}$ \\ \hline
$13,~31$& $4.6\times 10^{-5}$ & $4.6\times 10^{-5}$ \\ \hline
$22$& $9.0\times 10^{-3}$ & $0.012$ \\ \hline
$23,~32$& $2.2\times 10^{-4}$ & $2.2\times 10^{-4}$ \\ \hline
$33$& $0.018$ & $0.016$ \\ \hline \hline 
\end{tabular} 
\end{center}
\caption{{\it
The present upper limits on the products 
$\lambda \lambda^{\prime}$ relevant for 
the parton level reactions $d_j {\bar d}_k \to \sneu \to e^+ e^-$ 
and $d_j {\bar d}_k \to \sneut \to e^+ e^-$.  
$j,k=1$,2,3 are flavor indices and 
$n_i \equiv m_{{\tilde\nu}^i}/$[100 GeV] (see \protect\cite{foottables}).}}
\end{table}
\begin{table}
\begin{center}
\begin{tabular}{||r||r|r||}  \hline \hline 
\multicolumn{1}{||c||}{~~} & \multicolumn{2}{|c||}{~~} \\
\multicolumn{1}{||c||}{~~} & \multicolumn{2}{c||}{\raisebox{3mm}{$d_j {
\bar d}_k \to \sneui \to \mu^+ \mu^-$}}\\ \cline{2-3}
$jk \Downarrow$ & $(\sneue)~~\lambda_{212} 
\lambda^{\prime}_{1jk}/n_e^2$ & $(\sneut)~~\lambda_{232} 
\lambda^{\prime}_{3jk}/n_\tau^2$ \\ \hline \hline
$11$& $1.8\times 10^{-5}$ & $6.0\times 10^{-3}$ \\ \hline 
$12,~21$& $3.8\times 10^{-7}$  & $3.8\times 10^{-7}$ \\ \hline
$13,~31$& $2.4\times 10^{-5}$ & $2.4\times 10^{-5}$ \\ \hline
$22$& $1.0\times 10^{-3}$ & $0.012$ \\ \hline
$23,~32$& $5.5\times 10^{-5}$ & $5.5\times 10^{-5}$ \\ \hline
$33$& $3.5 \times 10^{-5}$ & $0.016$ \\ \hline \hline 
\end{tabular} 
\end{center}
\caption{{\it
The same as Table 1 for the parton level reactions 
$d_j {\bar d}_k \to \sneue \to \mu^+ \mu^-$ 
and $d_j {\bar d}_k \to \sneut \to \mu^+ \mu^-$
(see \protect\cite{foottables}).}}
\end{table}
\begin{table}
\begin{center}
\begin{tabular}{||r||r|r||}  \hline \hline 
\multicolumn{1}{||c||}{~~} & \multicolumn{2}{|c||}{~~} \\
\multicolumn{1}{||c||}{~~} & \multicolumn{2}{c||}{\raisebox{3mm}{$d_j {\bar d}_k \to 
\sneui \to \tau^+ \tau^-$}}\\ \cline{2-3}
$jk \Downarrow$ & $(\sneue)~~\lambda_{313} 
\lambda^{\prime}_{1jk}/n_e^2$ & $(\sneu)~~|\lambda_{323}| 
\lambda^{\prime}_{2jk}/n_\mu^2$ \\ \hline \hline
$11$& $1.1\times 10^{-6}$ & $5.4\times 10^{-3}$ \\ \hline 
$12,~13$& $6.0\times 10^{-5}$  & $5.4\times 10^{-3}$ \\ \hline
$21$& $1.1\times 10^{-4}$ & $0.011$ \\ \hline
$22$& $6.0\times 10^{-5}$ & $0.011$ \\ \hline
$23$& $6.0\times 10^{-4}$ & $0.011$ \\ \hline
$31$& $1.1\times 10^{-4}$ & $0.013$ \\ \hline
$32$& $9.9\times 10^{-4}$ & $0.022$ \\ \hline
$33$& $2.1 \times 10^{-6}$ & $0.022$ \\ \hline \hline 
\end{tabular} 
\end{center}
\caption{{\it
The same as Table 1 for the parton level reactions 
$d_j {\bar d}_k \to \sneue \to \tau^+ \tau^-$ 
and $d_j {\bar d}_k \to \sneu \to \tau^+ \tau^-$
(see \protect\cite{foottables}).}}
\end{table}  

We will focus here only on the $\mu \mu$ and
$\tau \tau$ production channels. 
In these cases our investigation simplifies since, as noted above, 
to a good approximation it is sufficient to consider an exchange of only
one  sneutrino flavor.  Production of $ee$ at the Tevatron for one sneutrino
flavor exchange was explored in
 \cite{zerwaspp}. There, only the $\hat s$-channel tau-sneutrino 
was investigated and only its coupling to the valence $d$-quark, 
{\it i.e.}, $\lambda^{\prime}_{311}$, was considered by assuming  
$\lambda^{\prime}_{3jk}=0$ for $j,k \neq 1$.      
Although, for the case of $ee$ production,  
the generalization to $\hat s$-channel exchanges of two 
sneutrino flavors is straightforward, we will not discuss it here. 

It was also shown in 
\cite{zerwaspp} that a potential discovery of sneutrino resonance 
formation in the Tevatron is possible  
through a study of the $\ell \ell$ invariant mass distribution   
$d \sigma^T/d M_{\ell \ell}$, wherein the sneutrino will unveil itself 
as a resonant peak.
However, in what follows we wish to take a different approach than that 
presented in \cite{zerwaspp}, in a further  attempt to constrain the $\rp$
couplings.  In particular, we will show that the 
study of total cross sections may also be very useful for obtaining limits 
on some of the $\rp$ couplings. 
For the purpose of bounding the $\rp$ couplings 
we will focus later on $\mu$-pair production. However, it should be 
emphasized at this point 
that for $\tau$-pair production, our total cross section analysis 
(with cuts on the $\tau\tau$ invariant mass) for constraining the $\rp$ 
couplings may prove to be more useful 
than a study of $d \sigma^T/d M_{\tau \tau}$. The reason is that it will 
be experimentally easier to determine the $\tau\tau$ production rate above 
some value of $M_{\tau \tau}$ than to reconstruct 
the $\tau \tau$ invariant mass on an event by event basis.        

\begin{figure}[htb]
\psfull
 \begin{center}
  \leavevmode
  \epsfig{file=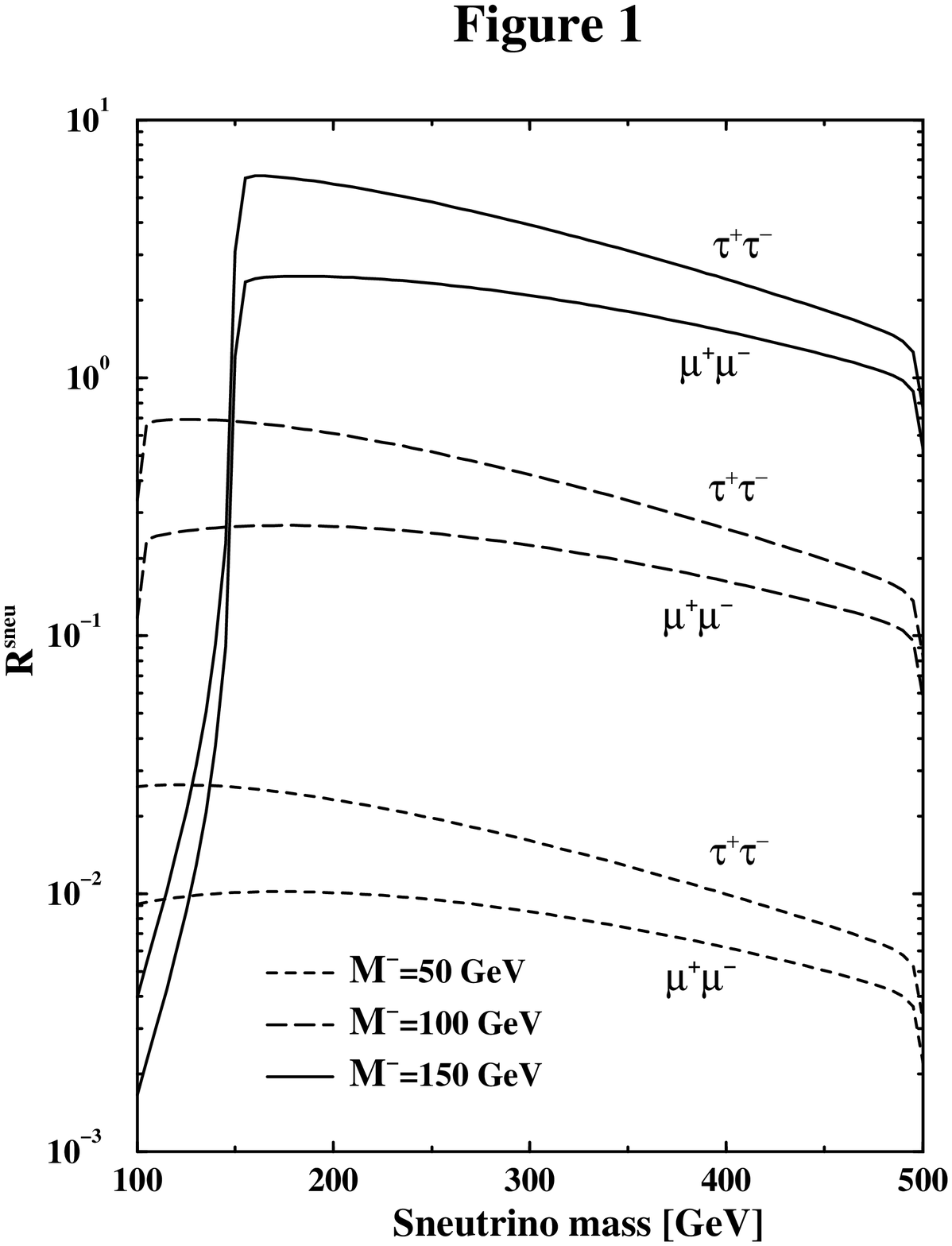,height=7cm,width=9cm,bbllx=0cm,bblly=2cm,bburx=20cm,bbury=25cm,angle=0}
 \end{center}
\caption{\emph{The ratio $R^{\rm sneu} \equiv
\sigma^T_{\sneupm^i}/\sigma^T_{SM}$, as  
a function of the $\tau$-sneutrino mass 
for the reaction $p \bar p \to \mu^+ \mu^- + X$ for: 
$M_{\mu \mu}^-=50$ GeV (lower dashed line), 
$M_{\mu \mu}^-=100$ GeV (lower long-dashed line), 
$M_{\mu \mu}^-=150$ GeV (lower solid line),
and as a function of the $\mu$-sneutrino mass 
for the reaction $p \bar p \to \tau^+ \tau^- + X$ for: 
$M_{\tau \tau}^-=50$ GeV (upper dashed line), 
$M_{\tau \tau}^-=100$ GeV (upper long-dashed line),
$M_{\tau \tau}^-=150$ GeV (upper solid line). 
We use $M_{\ell \ell}^+=500$ GeV, $\sqrt s=2$ TeV, 
$\Gamma_{\sneupm^i}=10^{-2}m_{\sneupm^i}$ and $m_{\sneup^i}=m_{\sneum^i}$.    
$M_{\ell \ell}^+$ and $M_{\ell \ell}^-$ are upper and lower cuts on 
the dilepton invariant mass, respectively.
Maximal allowed values of the 
products $\lambda \lambda^{\prime}$ are used 
for any given sneutrino mass on the horizontal axis,  {\it i.e.}, scaled as 
$m_{\sneui}/\left[100~{\rm GeV}\right]$ (see Tables 2 and 3). 
See also \cite{reffoot2}.}}
\label{fig1}
\end{figure}

To illustrate the magnitude of the sneutrino cross section as 
compared to the SM one, for $\mu \mu$ and $\tau \tau$ production, 
we plot in Figure 1 the ratio:

\bea
R^{\rm sneu} \equiv \frac{\sigma^T_{\sneui}}{\sigma^T_{SM}} ~,
\eea

\n where for the $\mu \mu$ case, $i=\tau$ and for $\tau \tau$ production, 
$i=\mu$. We show the ratio $R^{\rm sneu}$ for a $p \bar p$ c.m.\ energy of 
$\sqrt{s}=2$ TeV (which is assumed throughout the rest of the paper) 
and for three lower cuts on the $\ell \ell$ 
invariant mass, $M_{\ell \ell}^-=50,~100$ and 150 GeV, 
where we take $M_{\ell \ell}^+=500$ GeV\null.  
We include all combinations of $\left\{ j,k \right\}$ in $\lambda^{\prime}$,  
from valence and sea down-quarks (that is, for $\mu \mu$($\tau \tau$) 
production all combinations of 
$\lambda_{232} \lambda^{\prime}_{3jk}$($|\lambda_{323}|\lambda^{\prime}_{2jk}$)
are included). Also, for any $\left\{ j,k \right\}$
 we take the sneutrino mass-dependent 
maximal allowed values of the corresponding $\rp$ coupling, {\it i.e.}, scaled
as  $m_{\sneui}/\left[100~{\rm GeV}\right]$ (see Tables 2 and 3).  
Therefore, for any sneutrino mass value on the horizontal axis, 
Figure 1 represents the largest possible value of  
$R^{\rm sneu}$ that might be measured at $\sqrt s = 2$ TeV\null. 
As expected, with $M_{\ell \ell}^-=50$ GeV  where the SM $\hat s$-channel
$Z$  resonance dominates, we find that 
$R^{\rm sneu} \sim 10^{-2}$ for sneutrino masses between 100 and 500 GeV\null.
However, the sneutrino contribution is much more pronounced if 
the lower invariant mass cut is set to $M_{\ell \ell}^-=100$ or 150 GeV\null.
In the latter case, {\it i.e.,} $M_{\ell \ell}^-=150$ GeV, the   
$\hat s$-channel $Z$ resonance contribution 
is practically removed and it is possible  to have
$R^{\rm sneu} > 1$ over almost the entire 
sneutrino mass range 150--500 GeV\null. Therefore, these reactions 
may indeed lead to a discovery 
of a new scalar resonance, in particular, a sneutrino. 
For example, we see from Fig.~1 that with a lower cut 
of $M_{\ell \ell}^-=150$ GeV ($\ell=\tau~{\rm or}~\mu$), one can 
expect the total cross sections for $\tau^+\tau^-$ or $\mu^+ \mu^-$ 
pair production to increase by a factor of $\sim$ 50 or 20, respectively, 
if the sneutrino mass is $\sim 200$ GeV, {\it i.e.}, a spectacular discovery. 

Let us now illustrate how a study of 
total cross sections for $p \bar p \to \ell^+ \ell^- +X$ can be used to 
constrain some of the $\rp$ couplings. Even in the most general 
case in which all combinations of the products $\lambda \lambda^{\prime}$ 
are taken into account, one can take advantage of the information given 
in Tables 1--3 combined with the corresponding 
numerical values of the $c_{jk}$'s (defined in Eq.~\ref{cjk})
to greatly simplify the analysis. 
The simplest example perhaps is the reaction 
$p \bar p \to \mu^+ \mu^- +X$ which is also the easiest to measure. 
In $\mu$-pair production, as already mentioned above, to a good 
approximation only $\sneut$ 
with flavor diagonal couplings to $d \bar d$, $s \bar s$ and $b \bar b$ 
contributes. Moreover, owing to 
the very small probability of finding a $b$-quark in the 
proton with momentum fraction $x_b$, 
$b \bar b$ annihilation is negligible compared to the valence $d \bar d$ and 
sea $s \bar s$ fusion. We are therefore left with only two relevant $\rp$ 
couplings: $\sqrt {v_{\tau 2}^{11}}=\lambda_{232} \lambda^{\prime}_{311}$ 
and $\sqrt {v_{\tau 2}^{22}}=\lambda_{232} \lambda^{\prime}_{322}$. 
In fact, although the present limits allow 
$v_{\tau 2}^{22} = 4 v_{\tau 2}^{11}$ (see Table 3), we still find that 
the valence $d \bar d$ contribution dominates over  
the sea $s \bar s$ annihilation due to larger probability 
distributions (calculated from Eq.~\ref{cjk}). 
Nonetheless, for completeness we 
include in our analysis below both the $d \bar d$ and the $s \bar s$ 
contributions.         
   
\begin{figure}[htb]
\psfull
 \begin{center}
  \leavevmode
  \epsfig{file=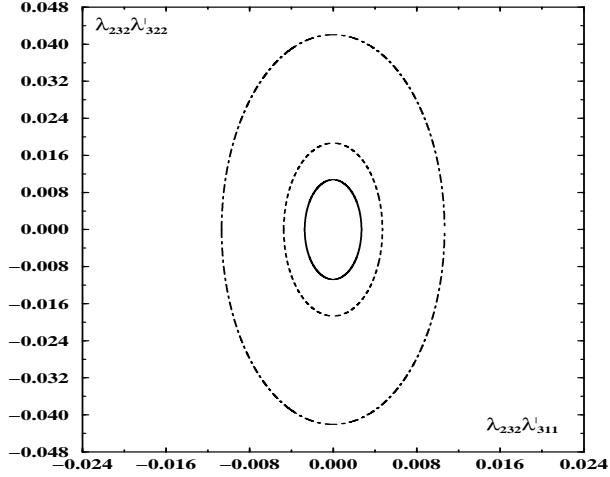,height=7cm,width=9cm,bbllx=0cm,bblly=2cm,bburx=20cm,bbury=25cm,angle=0}
 \end{center}
\caption{\emph{The attainable $1\sigma$-limit contours 
on the $\rp$ coupling products $\lambda_{232} \lambda_{311}^{\prime}$ 
and $\lambda_{232} \lambda_{322}^{\prime}$, for the reaction 
$p \bar p \to \mu^+ \mu^- + X$ 
at the Tevatron Run II with c.m.\ energy $\sqrt s=2$ TeV
and an integrated luminosity of ${\cal L}=2$ fb$^{-1}$. 
For illustration we choose $m_{\sneup^\tau}=m_{\sneum^\tau}=200$ GeV and 
take $M_{\mu \mu}^+=500$ GeV\null. We use: 
$M_{\mu \mu}^-=50$ GeV (dotted-dashed line) \cite{footsystematics},
$M_{\mu \mu}^-=100$ GeV (dashed line) and
 $M_{\mu \mu}^-=150$ GeV (solid line). See also caption to Figure 1.}}
\label{fig2}
\end{figure}

The statistical significance of the deviation of the total cross section 
from the SM cross section is defined by:

\bea
N_{SD} = \frac{|\sigma^T - \sigma_{SM}^T| {\cal L}}
{\sqrt {\sigma^T {\cal L}}}~,
\eea 

\n which in our case, {\it i.e.}, 
$\sigma^T = \sigma^T_{\sneui} + \sigma^T_{SM}$ and $i=\tau$, $p=2$ 
(see Eq.~\ref{sigsneu0}), 
simplifies to:

\bea
N_{SD}= \frac{\sum_{jk} c_{jk} v_{\tau 2}^{jk}}{\sqrt {\sum_{jk} c_{jk} 
v_{\tau 2}^{jk} +\sigma_{SM}^T}} \sqrt {\cal L} 
~~\stackrel{d \bar d,s \bar s~{\rm only}}
{\longrightarrow} ~~ \frac{c_{11} v_{\tau 2}^{11} + 
c_{22} v_{\tau 2}^{22}}{\sqrt {c_{11} v_{\tau 2}^{11}+
c_{22} v_{\tau 2}^{22} +\sigma_{SM}^T}} \sqrt {\cal L} \label{nsd}~,
\eea  

\n where ${\cal L}$ is the integrated luminosity at the Tevatron. 
Using Eq.~\ref{nsd} we can now plot contours of $N_{SD}$ limits 
on the two $\rp$ couplings 
$\sqrt {v_{\tau 2}^{11}}=\lambda_{232} \lambda^{\prime}_{311}$ 
and $\sqrt {v_{\tau 2}^{22}}=\lambda_{232} \lambda^{\prime}_{322}$. 
In Figure 2 we show the $1 \sigma$-limit 
contours corresponding to the three lower
$\mu \mu$ invariant mass cuts $M_{\mu \mu}^-=50,~100$ and 150 GeV, again 
taking $M_{\mu \mu}^+=500$ GeV\null. For illustration we choose 
$m_{\sneut_{\pm}}=200$ GeV 
and a total integrated luminosity of ${\cal L}=2$ fb$^{-1}$ 
appropriate for the Tevatron Run II\null. The values at the 
end points of the $x$ and $y$ axes in Figure 2 are the presently allowed 
upper limits on the coupling products $\lambda_{232} \lambda^{\prime}_{311}$  
and $\lambda_{232} \lambda^{\prime}_{322}$, respectively, 
for $m_{\sneupm^\tau}=200$ GeV\null.               
We observe that, due to larger valence $d$-quark probability functions,  
a significant improvement over the present limits may be 
obtained for the $\rp$ product $\lambda_{232} \lambda^{\prime}_{311}$ 
\cite{footsystematics}. As expected, the effect is much more pronounced 
when the SM ``background'' is removed, {\it i.e.}, $M_{\mu \mu}^-=150$ GeV, 
where the attainable $1\sigma$ limits are: 
$ -0.003 \lsim \lambda_{232} \lambda^{\prime}_{311} \lsim 0.003$ and     
$ -0.011 \lsim \lambda_{232} \lambda^{\prime}_{322} \lsim 0.011$, for a 
$\tau$-sneutrino mass 
of 200 GeV. For $\lambda_{232} \lambda^{\prime}_{311}$ this limit 
is about one order of magnitude better than the current limit and for 
$\lambda_{232} \lambda^{\prime}_{322}$ it provides an improvement 
by about a factor of 4 over the present limit.
\\

\n \underline{\bf 3. Sneutrino mixing and 
\boldmath ${\tau}$-polarization asymmetries in 
\boldmath ${p \bar p \to \tau^+ \tau^- +X}$}\\

\n In this section we explore the possibility of studying detailed 
properties of the sneutrino sector in the MSSM with $\rp$. 
In particular, we discuss the reaction
$p \bar p \to \tau^+ \tau^- +X$ proceeding via muon-sneutrino exchange.
Therefore, throughout the rest of the section 
we drop the sneutrino index $\mu$ 
and denote $m_{\sneupm^\mu} \equiv m_{\pm}$, $\Gamma_{\sneup^\mu} 
\simeq \Gamma_{\sneum^\mu} \equiv \Gamma$ and set  
$\Gamma=10^{-2}m_-$.   We focus on two issues:

\begin{enumerate}
\item Detection of muon-sneutrino mixing, {\it i.e.}, 
probing a possible mass splitting between the muon-sneutrino 
CP-even ($\sneup$) and CP-odd ($\sneum$) states.

\item The possibility of having large CP-violating signals 
at the Tevatron, which, in the presence of the $\rp$ $\hat s$-channel 
$\sneupm$ exchanges, are generated already at the tree-level  
if there is a non-vanishing mass splitting 
$\Delta m \equiv m_{+} -  m_- \neq 0$.
\end{enumerate}

\n As demonstrated below, these two effects can be studied 
at the Tevatron through measurements of some specific CP-violating 
and CP-conserving $\tau$-double-spin asymmetries which were suggested in 
\cite{hepph9802251} for the reaction $\ell^+ \ell^- 
\to \tau^+ \tau^-$ appropriate 
to leptonic colliders. Note that spin asymmetries in 
$p \bar p \to \ell^- \ell^+ +X$ can be measured only for $\ell=\tau$; 
the electron or the muon polarization is not accessible in such high energy
collider experiments.

The CP-violating and CP-conserving  
$\tau$-double-spin asymmetries are derived at the parton level, 
{\it i.e.}, $q_j {\bar q}_k \to \tau^+ \tau^-$, following  the same line 
of arguments as introduced in our previous work \cite{hepph9802251} and are 
then ``dressed'' with the parton distribution functions.
In the rest frame of $\tau^-$ we define the basis vectors: 
$\vec{e}_z \propto - ({\vec{p}}_{q_j} + {\vec{p}}_{{\bar q}_k})$, 
$\vec{e}_y \propto 
{\vec{p}}_{q_j} \times {\vec{p}}_{{\bar q}_k}$ and     
$\vec{e}_x = \vec{e}_y \times \vec{e}_z$. 
For the $\tau^+$ we use a similar set of definitions such that 
$\vec{{\bar e}}_x, \vec{{\bar e}}_y, \vec{{\bar e}}_z$ 
are related to $\vec{e}_x,\vec{e}_y,\vec{e}_z$ by charge conjugation.
 We then introduce the following $\tau^+\tau^-$ double-polarization 
operator with respect to each of the coordinate directions defined above:

\bea
{\hat \Pi}_{mn} \equiv \frac{ 
N({\bar {\uparrow}}_m \uparrow_n) - 
N({\bar {\uparrow}}_m \downarrow_n) - 
N({\bar {\downarrow}}_m \uparrow_n) +
N({\bar {\downarrow}}_m \downarrow_n) }
{ N({\bar {\uparrow}}_m \uparrow_n) + 
N({\bar {\uparrow}}_m \downarrow_n) + 
N({\bar {\downarrow}}_m \uparrow_n) +
N({\bar {\downarrow}}_m \downarrow_n)} 
~, \label{piij} 
\eea  

\n where $m,n=x,y,z$. For example, $N({\bar {\uparrow}}_x \uparrow_y)$
stands for the number of events in which $\tau^+$ has 
spin +1 in the direction $x$ in its rest frame 
and $\tau^-$ has spin +1 in the direction $y$ in its 
rest frame. The spin vectors of $\tau^+$ and $\tau^-$ 
are therefore defined in their respective rest frames as: 
${\vec s}\,^+= ({\bar s}_x,{\bar s}_y,{\bar s}_z)$ and ${\vec s}\,^-= 
(s_x,s_y,s_z)$ and  ${\hat \Pi}_{mn}$ is calculated at the parton level 
in the $q_j {\bar q}_k$ c.m.\ frame by boosting 
${\vec s}\,^+$ and ${\vec s}\,^-$ from the $\tau^+$ and $\tau^-$ rest 
frames to the $q_j {\bar q}_k$ c.m.\ frame.

It is easy to verify that the ${\hat \Pi}_{mn}$'s possess the 
following transformation properties under the operations 
of CP and that of the naive time reversal $T_N$ \cite{foot3}:

\bea
&&{\rm CP}({\hat \Pi}_{mn})={\hat \Pi}_{nm}~~~~~~ {\rm for~ all}~ m,n~,\\ 
&&T_N({\hat \Pi}_{mn})=-{\hat \Pi}_{mn}~~~~ {\rm for}~m~{\rm or}~n=y~
{\rm and}~ m \neq n~,\\
&&T_N({\hat \Pi}_{mn})={\hat \Pi}_{mn}~~~~~~ {\rm for}~m,n \neq y~
{\rm and~ for}~ m=n=x,y,z~.
\eea
 
\n We can therefore define:

\bea
{\hat A}_{mn}=\frac{1}{2} \left({\hat \Pi}_{mn}-{\hat \Pi}_{nm} \right) ~~,~~
{\hat B}_{mn}=\frac{1}{2} \left({\hat \Pi}_{mn}+{\hat \Pi}_{nm} \right)
~,\label{asym}
\eea

\n such that ${\hat A}_{mn}$ are CP-odd (${\hat A}_{mm}=0$ by definition) 
and ${\hat B}_{mn}$ are CP-even. Also, ${\hat A}_{xy},{\hat A}_{zy},{\hat
B}_{xy}$ and ${\hat B}_{zy}$ are 
$T_N$-odd while ${\hat A}_{xz},{\hat B}_{xz},{\hat B}_{xx},{\hat B}_{yy}$
and ${\hat B}_{zz}$ are $T_N$-even.    

In the SM, as expected, there is no CP violation at the tree-level and we 
find that  only the CP-even asymmetries ${\hat B}_{xx}, {\hat B}_{yy}$
and ${\hat B}_{zz}$ are non-zero at the tree-level, where in fact, 
 for either $u \bar u$ or $d \bar d$ annihilation,  
${\hat B}_{xx}=-{\hat B}_{yy}$ and ${\hat B}_{zz}=1$ \cite{hepph9802251}. 
However, in the $\hat s$-channel $\sneupm$ exchange case, 
the CP-violating asymmetry ${\hat A}_{xy}$, being a $T_N$-odd quantity,  
is also non-zero already at the tree-level. In particular, 
for any given flavor combination $\left\{j,k \right\}$ in 
$d_j {\bar d}_k \to \sneupm \to \tau^+ \tau^-$,
${\hat B}_{xx}, {\hat B}_{yy}, {\hat B}_{zz}$ and ${\hat A}_{xy}$    
are given by \cite{hepph9802251} 
(recall that $\lambda_{323} = (a+ib)/\sqrt 2$):

\bea
{\hat A}_{xy}=\left(\frac{2a b}{a^2+b^2}\right) 
\frac{D_-}{D_+}~~,~~{\hat B}_{xx}={\hat B}_{yy}=
\left(\frac{a^2-b^2}{a^2+b^2}\right) \frac{D_-}{D_+}
~~,~~{\hat B}_{zz}=-1~, \label{sneuasym}
\eea  

\n where:

\bea
D_\pm \equiv |\pi_+|^2 \pm |\pi_-|^2~~,~~ \pi_\pm = 
\left(s -m_{\pm}^2 +i m_{\pm} \Gamma \right)^{-1} 
\label{dprop1}~.
\eea
 
\noindent While the presence of a non-vanishing tree-level CP-nonconserving 
asymmetry in $p \bar p \to \tau^+ \tau^- +X$ is a unique
outcome of sneutrino resonance formation, 
as was mentioned above, the CP-even spin-asymmetries receive 
contributions from the SM too.   However, it is also possible to 
construct a CP-even $\tau$-spin asymmetry which is sensitive  
only to the $\hat s$-channel $\sneupm$ exchanges and is identically 
zero in the SM\null. This CP-conserving asymmetry is defined as    
\cite{hepph9802251}, ${\hat B} \equiv ({\hat B}_{xx}+{\hat B}_{yy})/2$. 
Obviously, at tree-level, 
${\hat B}=0$ in the SM (due to ${\hat B}_{xx}=-{\hat B}_{yy}$)
 and ${\hat B}={\hat B}_{xx}={\hat B}_{yy}$ for the sneutrino case. 
Thus, a measurement of ${\hat B}$ substantially different
from zero will be a strong indication for the existence of new physics in 
$p \bar p \to \tau^+ \tau^- +X$, in the form of  
new non-vanishing $\hat s$-channel scalar exchanges and        
will provide explicit information 
on the new $\tau \sneupm^{\mu} \tau$ coupling. 
The use of $\tau$-spin asymmetries as a probe of new physics was also 
suggested in \cite{bern,hagiwara}, {\it e.g.,} in the decay $H^0 \to 
\tau^+ \tau^-$.  
Note that the observable ${\hat B}_{zz}$, which in the SM simply 
translates to the spin correlation 
$<{\vec s}\,^+ \cdot {\vec s}\,^->$ that was suggested in \cite{bern} for 
the decay $H^0 \to \tau^+ \tau^-$,  may also be useful in 
distinguishing between the SM vector-boson exchanges and the sneutrino 
scalar exchanges. However, since 
${\hat B}_{zz}=-1$ for the pure $\sneupm$ exchange case,   
a measurement of ${\hat B}_{zz}$, will be insensitive 
to the couplings $a$ and $b$ in $\lambda_{323}$.       

In what follows, we will therefore discuss only the  
CP-odd ${\hat A}_{xy}$ and the CP-even ${\hat B}$ spin-asymmetries. 
Let us comment first on how to measure these spin-asymmetries. 
Since the $\tau$ spins are not directly observable, the $\tau$ decay products 
should be used as spin analyzers (see \cite{bern,hagiwara,foottaupol,tsay}). 
In particular,
following \cite{bern}, since 
in the $\tau^- \tau^+$ c.m. frame
our asymmetry ${\hat A}_{xy}$ corresponds to  
the spin correlation 
${\hat p}_{\tau^-} \cdot ({\vec s}\,^- \times {\vec s}\,^+)$ 
 ({\it i.e.,} the operator ${\cal O}_2$ 
in \cite{bern}), in the case of the 
two-body decays $\tau^\pm \to \pi^\pm {\bar \nu}_\tau (\nu_\tau)$,  
${\hat A}_{xy}$ can be translated to the correlation 
${\hat p}_{\tau^-} \cdot ({\hat p}^\ast_{\pi^+} \times 
{\hat p}^\ast_{\pi^-})$, where ${\hat p}_{\tau^-}$ is the flight direction 
of $\tau^-$ in the $\tau^- \tau^+$ c.m. frame and ${\hat p}^\ast_{\pi^\pm}$ 
is the flight direction of $\pi^\pm$ in the corresponding 
$\tau^\pm$ rest frames, respectively. Thus, in order to measure      
${\hat A}_{xy}$ one can trace the $\tau^\pm$ spins by measuring 
the flight direction of their decay products in their respective rest frames 
or equivalently by measuring the triple correlation 
$\langle {\hat p}_{\tau^-} \cdot ({\hat p}^\ast_{\pi^+} \times 
{\hat p}^\ast_{\pi^-}) \rangle$. Note that similar correlations 
between the $\tau^\pm$ 
spins and their decay products can be obtained for the other combinations 
of the $\tau^\pm$ two-body decays (see Table II in \cite{bern}).    
In the same way, ${\hat B}$ is proportional to spin correlation 
${\vec s}\,^- \cdot {\vec s}\,^+ - ({\hat p}_{\tau^-} \cdot {\vec s}\,^-) 
({\hat p}_{\tau^-} \cdot {\vec s}\,^+)$ 
(this corresponds to the operator $({\cal O}_3 - {\cal O}_4)$ 
in \cite{bern})
and can be translated to a correlation between the $\tau^\pm$ decay 
products, {\it e.g.,} in the case of $\tau^\pm \to \pi^\pm \nu_{\tau^\pm}$,  
$\langle {\cal O}_3 - {\cal O}_4 \rangle \propto 
\langle {\hat p}^\ast_{\pi^+} \cdot {\hat p}^\ast_{\pi^-} \rangle 
- \langle ({\hat p}_{\tau^-} \cdot {\hat p}^\ast_{\pi^-}) 
({\hat p}_{\tau^-} \cdot {\hat p}^\ast_{\pi^+}) \rangle$ \cite{bern}.

It is important to note that these two asymmetries change 
sign around $\sqrt {\hat s} \sim m_-$ (see
\cite{hepph9802251}). In fact, roughly 
${\hat A}_{xy} (\sqrt {\hat s}=m_- - \delta m) \sim -  
{\hat A}_{xy} (\sqrt {\hat s}=m_- + \delta m)$ and similarly for ${\hat B}$,
for a given mass shift $\delta m$. 
Of course, this does not introduce any difficulty in a leptonic collider 
where the c.m.\ energy of the colliding leptons is fixed. 
However, for the Tevatron, after folding in the parton luminosities 
(see Eqs.~\ref{sigpp} and \ref{partonlum}) and  
integrating over $\sqrt {\hat s}$ 
(or equivalently $M_{\tau \tau}$), due to this change in sign, the effects
 become too small to be of any measurable 
consequences. To bypass this problem we suggest here a two-step measurement;
at the first stage one would have to identify the sneutrino resonant peak 
by measuring the $\tau \tau$ invariant mass distribution 
$d \sigma^T / dM_{\tau \tau}$. Then, once 
the sneutrino mass is known, the asymmetries ${\hat A}_{xy}$ and 
${\hat B}$ may be multiplied by the sign of $(M_{\tau \tau} - m_-)$ 
to account for the relative minus sign as one goes from 
$M_{\tau \tau} < m_-$ to $M_{\tau \tau} > m_-$:

\begin{figure}[htb]
\psfull
 \begin{center}
  \leavevmode
  \epsfig{file=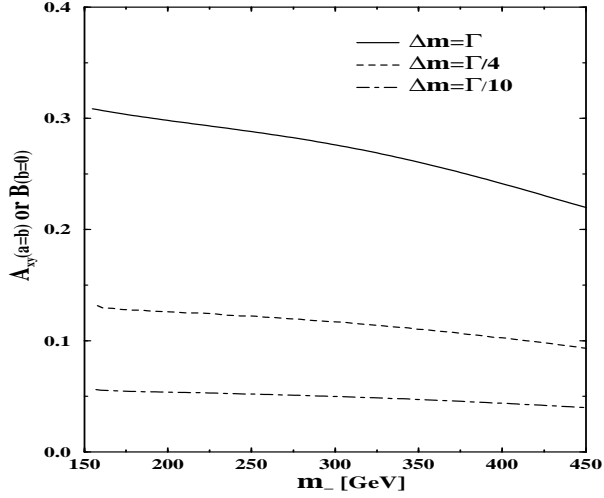,height=7cm,width=9cm,bbllx=0cm,bblly=2cm,bburx=20cm,bbury=25cm,angle=0}
 \end{center}
\caption{\emph{The maximal values of $A_{xy}$ and $B$   
as a function of the lighter CP-odd muon-sneutrino mass $m_-$, for 
three mass-splitting values $\Delta m=\Gamma$ (solid line), 
$\Delta m=\Gamma/4$ (dashed line) and 
$\Delta m=\Gamma/10$ (dotted-dashed line).
 $M_{\tau \tau}^-=150$ GeV, $M_{\tau \tau}^+=500$ GeV, 
$\sqrt s=2$ TeV and $\Gamma=10^{-2}m_-$.}}
\label{fig3}
\end{figure}

\bea
{\hat A}_{xy} \longrightarrow {\rm sgn}(M_{\tau \tau} - m_-)  
{\hat A}_{xy}~~,~~{\hat B} \longrightarrow 
{\rm sgn}(M_{\tau \tau} - m_-)  
{\hat B}~.
\eea

\n The corresponding CP-odd and CP-even asymmetries for the overall 
reaction $p \bar p \to \tau^+ \tau^- +X$,  $A_{xy}$ and $B$, 
are then given by:

\begin{figure}[htb]
\psfull
 \begin{center}
  \leavevmode
  \epsfig{file=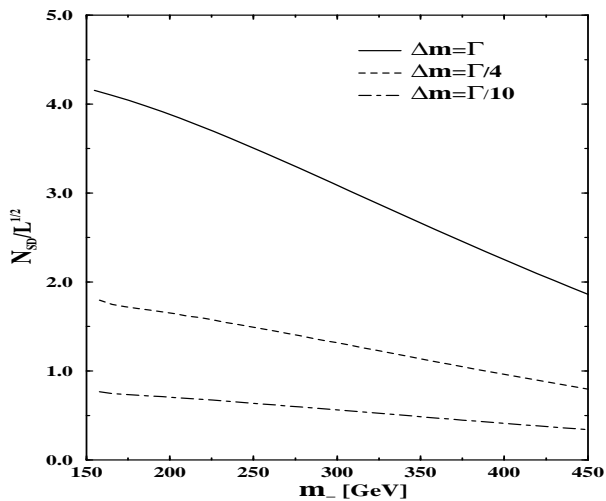,height=7cm,width=9cm,bbllx=0cm,bblly=2cm,bburx=20cm,bbury=25cm,angle=0}
 \end{center}
\caption{\emph{The scaled statistical significance, $N_{SD}/\sqrt L$, 
for  both $A_{xy}$ and $B$, as a function of $m_-$. As in Figure 3, 
$A_{xy}$ and $B$ are
evaluated  at their maximal values. For the Tevatron runs II, III the 
statistical significance, $N_{SD}$, is obtained 
by multiplying the values on the $y$ axis by 
$\sqrt L=\sqrt 2$, $\sqrt L=\sqrt {30}$, respectively. 
See also caption to Figure 3.}}
\label{fig4}
\end{figure}

\bea
A_{xy};~B = \frac{\sum_{j,k} \int_{\tau_-}^{\tau_+} 
d \tau ~ \frac{d {\cal L}_{jk}(\tau)}
{d \tau} ~ {\hat \sigma}^{jk}_{\sneupm}
 (\hat s =\tau s) \cdot {\rm sgn}(\sqrt {\hat s} - m_-) \cdot 
{\hat A}_{xy} ;~ {\hat B}}
{\sigma^T_{\sneupm} + \sigma^T_{SM}}
\label{ppasym}~.
\eea     
 
\noindent From Eq.~\ref{sneuasym} we 
observe that both ${\hat A}_{xy}~{\rm and}~{\hat B} \propto D_-/D_+$,  
where the proportionality 
factors do not depend on the absolute magnitude of the couplings 
$a$ and $b$ in $\lambda_{323}$ but rather on any function of their 
ratio $f(a/b)$. 
As in \cite{hepph9802251}, without loss of generality, we will assume that 
$a$ and $b$ are positive and study the 
asymmetries as a function of the ratio $r \equiv b/(a+b)$. 
Thus $r$ can vary between $0 \leq r \leq 1$, 
where the lower and upper limits of $r$ are given by $b=0$ and 
$a=0$, respectively. One can immediately observe that ${\hat A}_{xy}$ 
and ${\hat B}$ complement each other as they probe opposite ranges of $r$.
For ${\hat A}_{xy}$ the maximal value $D_-/D_+$ is obtained when 
$r=1/2$ ($a=b$) and the largest positive $\hat B$ possible is 
${\hat B}=D_-/D_+$ when $r=0$ ($b=0$). Also, at 
$r=1$ ($a=0$), ${\hat B}=-D_-/D_+$, 
thus reaching its maximum negative value.

In Figure~3 we plot the maximal values of $A_{xy}$ and $B$, 
corresponding to the maxima 
${\hat A}_{xy}={\hat B}=D_-/D_+$, as a function 
of the lighter muon-sneutrino mass $m_-$.   
We take $M_{\tau \tau}^-=150$ GeV, $M_{\tau \tau}^+=500$ GeV and 
the mass splitting values $\Delta m =\Gamma,~
\Gamma /4,~\Gamma /10$ (recall that $\Gamma = 10^{-2} m_{-}$). 
Also, for completeness we include all $\left\{ j,k \right\}$ 
flavor combinations of the annihilating down quarks with their 
corresponding $\rp$ couplings (see Table~3). 
Evidently, $A_{xy}$ and $B$ can reach $\sim 20$--30\% 
throughout the entire mass range, $150~{\rm GeV} \lsim m_- \lsim 450$
GeV, if $\Delta m =\Gamma$, and $\sim 10 - 13\%$ even with a 
smaller splitting of $\Delta m = \Gamma /4$.  
  
The statistical significance, $N_{SD}$, 
with which $A_{xy}~{\rm or}~B$ can be detected at the Tevatron, is given 
by $N_{SD} = \sqrt N |{\cal A}| \sqrt {\epsilon}$ \cite{foottaupol}, where 
${\cal A}=A_{xy}~{\rm or}~B$ are given in Eq.~\ref{ppasym}, 
$N=(\sigma^T_{\sneupm}+\sigma^T_{SM}) \times {\cal L}$ 
is the total number  of $p \bar p \to \tau^+ \tau^- +X$ events and 
${\cal L}$ is the integrated luminosity at the Tevatron.
$\epsilon$ is the combined efficiency for the simultaneous measurement
of  the $\tau^+$ and $\tau^-$ spins which, therefore,   depends on the
efficiency for the spin analysis  and also on the branching ratios 
of the specific $\tau^+$ and $\tau^-$ decay channels that are being analyzed. 
The simplest examples perhaps are the two-body 
decays $\tau^\pm \to \pi^\pm \nu_\tau$ and $\tau^\pm \to \rho^\pm \nu_\tau$, 
although 3-body decays may also be useful 
\cite{bern,hagiwara,tsay}. 
For these two-body decays the decay density matrix 
is \cite{bern,hagiwara,tsay} 
$\left( 1 \mp \alpha_X {\hat p}_{X^\pm}^\ast \cdot 
{\vec s}\,^{\pm} \right) d \Omega_{X^\pm}/4 \pi$,
where $X^\pm = \pi^\pm$ or $\rho^\pm$ and $\alpha_X$ is the spin analyzing 
quantity which determines the sensitivity for measuring the $\tau$-spin via 
the momentum, ${\hat p}_{X^\pm}^\ast$, of its decay product. 
In particular, $\alpha_\pi=1$ and $\alpha_\rho=0.456$ 
\cite{bern,hagiwara,tsay}. 
Now, since in practice one measures the momenta of the $\tau$ decay products 
in order to trace its spin, to determine the expected statistical 
significance one has to multiply 
the asymmetries $A_{xy}$ and $B$ by the  
quality of the spin analysis in 
a given decay scenario of the $\tau^- \tau^+$ pair.
Thus, for example, if  
$\tau^- \to \rho^- \nu_\tau$ and $\tau^+ \to \pi^+ {\bar \nu}_\tau$, 
$A_{xy}$ and $B$ are suppressed by $\alpha_\rho \times \alpha_\pi = 0.456$ 
(and $|A_{xy}|^2$ and $|B|^2$ are suppressed by        
 $(\alpha_\rho \times \alpha_\pi)^2 = 0.456^2)$.       
Therefore, when all combinations of the above $\tau^+,\tau^-$ 
two-body decay channels are taken into account 
the combined efficiency is given by 
$\epsilon = \sum_{X_1,X_2} {\rm Br}_{X_1} {\rm Br}_{X_2} \times 
( \alpha_{X_1} \alpha_{X_2} )^2 \sim 0.03$, 
where ${\rm Br}_{X_i}$ is the branching ratio 
for $\tau$ to decay to $X_i = \pi$ or $\rho$, \cite{bern}.
We will adopt this number henceforward \cite{foottaupol}.

\begin{figure}[htb]
\psfull
 \begin{center}
  \leavevmode
  \epsfig{file=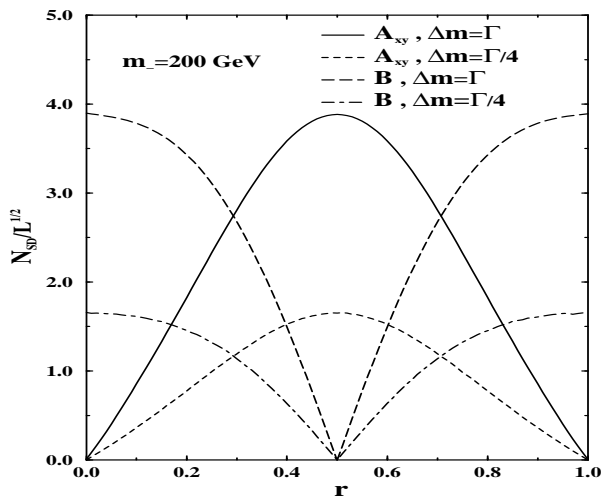,height=7cm,width=9cm,bbllx=0cm,bblly=2cm,bburx=20cm,bbury=25cm,angle=0}
 \end{center}
\caption{\emph{$N_{SD}/\sqrt L$ for $A_{xy}$ and $B$ 
as a function of $r \equiv b/(a+b)$. The cases $\Delta m= \Gamma$ and 
$\Delta=\Gamma/4$ are illustrated. See also captions to Figures 3 and 4.}}
\label{fig5}
\end{figure}

In Figure~4 we scale out the luminosity factor from the theoretical
prediction, by  plotting $N_{SD}/\sqrt L$ --- where $L \equiv {\cal L}/$(1
fb $^{-1}$) has no units --- corresponding to $A_{xy}$ and $B$ 
at their maximal values at the upgraded Tevatron with $\sqrt s=2$ TeV,
 as a function of $m_-$.
We choose the same values for $\Delta m$ and for $M_{\tau \tau}^\pm$ 
as in Figure 3 and we again include all $\left\{ j,k \right\}$ 
flavor combinations in $d_j {\bar d}_k$ fusion.  
We see that with $\Delta m =\Gamma$,
$4.2 \lsim N_{SD}/\sqrt L \lsim 1.8$ within the mass range
$155~{\rm GeV} \lsim m_- \lsim 450$ GeV\null. Also,  
$1.8 \lsim N_{SD}/\sqrt L \lsim 0.8$ for $\Delta m =\Gamma/4$,
within the same muon-sneutrino mass range. 
Thus, for example, at the Tevatron Run II with ${\cal L}=2$ fb$^{-1}$ 
\cite{tevatron} 
both the CP-violating and the CP-conserving spin asymmetries may be 
detected with a sensitivity above $3\sigma$  over the mass 
range $155~{\rm GeV} \lsim m_- \lsim 400$ GeV if 
$\Delta m =\Gamma$. These asymmetries are detectable, 
at the Tevatron Run III with ${\cal L}=30$ fb$^{-1}$ \cite{tevatron}, 
with a statistical significance above $3\sigma$ over the entire mass 
range $155~{\rm GeV} \lsim m_- \lsim 450~{\rm GeV}$
for $\Delta m =\Gamma/4$. In fact, for this case,  
even if the splitting is as small as $\Delta m =\Gamma/10$, 
a $3 \sigma$ detection of the CP-odd and CP-even spin asymmetries 
is feasible in Run III within the $\sim 150$ GeV mass range 
$155~{\rm GeV} \lsim m_- \lsim 300$ GeV\null.

Finally, Figure~5 shows the dependence of $N_{SD}/\sqrt L$, 
for $A_{xy}$ and $B$,  on the ratio $r$, where as in Figures~3 and 4, 
all $\left\{ j,k \right\}$ flavor combinations are included 
and $M_{\tau \tau}^-=150$, $M_{\tau \tau}^+=500$ GeV, respectively.  
For illustration we set $m_-=200$ GeV and chose 
$\Delta m =\Gamma$ and $\Delta m =\Gamma/4$ for both 
 $A_{xy}$ and $B$. Recall that for the Tevatron Runs II, III the 
statistical significance, $N_{SD}$, is obtained 
by multiplying the values on the $y$ axis by 
$\sqrt L=\sqrt 2$, $\sqrt L=\sqrt {30}$, respectively. 
We therefore observe  that a simultaneous measurement of
$A_{xy}$ and $B$ can cover practically 
the entire range of $r$ with a statistical sensitivity 
above $3 \sigma$ at the Tevatron Run II, if $\Delta m =\Gamma$.
At the Tevatron Run III, the entire range of $r$ is covered with at least 
$3\sigma$ standard deviations even for $\Delta m =\Gamma/4$.\\

\n \underline{\bf 4. Summary and conclusions}\\

\n To summarize, we found that if the data do not deviate from the SM, 
then existing limits on some products of 
the $\rp$ couplings $\lambda$ and $\lambda^{\prime}$ can be substantially 
improved in future runs of the Tevatron through a study 
of total cross sections for the reactions $p \bar p \to \ell^+ \ell^- +X$.
As an example, we have considered $\mu^+\mu^-$ pair production channel 
and found that at the Tevatron Run II the existing limits on the 
two specific $\rp$ coupling products $\lambda_{232} \lambda^{\prime}_{311}$
and  $\lambda_{232} \lambda^{\prime}_{322}$ 
can be improved by  a factor of $\sim 10$ and $\sim 4$, respectively.

We have also introduced CP-violating and CP-conserving $\tau$-double-spin
asymmetries  and applied them to the process $p \bar p \to \tau^+ \tau^-
+X$.  We have shown that two of these spin asymmetries are unique in 
their ability to distinguish between the CP-odd and CP-even muon-sneutrino 
mass eigenstates in $p \bar p \to \sneupm^\mu +X \to \tau^+ \tau^- +X$. 
Both asymmetries arise already at the tree-level and can become large, 
of the order of tens of a percent. 
Although the $\tau$ spins have not been measured at the Tevatron yet,  
the large CP-odd and CP-even effects that are possible  
in the future runs of the Tevatron may motivate the experimentalists 
for a serious attack on the problem.   
Only after an experimental setup for measuring the $\tau$ spins 
is established will the spin asymmetries suggested in this work become  
observable with a sensitivity above $3\sigma$ 
at future runs of the Tevatron provided that the efficiencies for
detecting the $\tau$ spin approach $\sim 10\%$. 
In such an optimistic scenario, the above $3\sigma$ signal is possible
 throughout almost the entire muon-sneutrino mass range 
$150~{\rm GeV} \lsim m_{\sneupm^\mu} \lsim 450$ GeV\null.
We have shown that such significant CP-violating and CP-conserving
 signals can arise 
even for a mass splitting between the CP-odd and CP-even $\sneupm^\mu$ 
states below their corresponding widths, {\it i.e.}, $\Delta m < \Gamma$. 
They may therefore serve as extremely powerful probes of the sneutrino mixing 
phenomenon.

As far as CP-violation is concerned, it is especially 
gratifying that such a large CP-nonconserving effect may arise in 
$\tau$-pair production  at the Tevatron  
and may be searched for in the near future.

\bigskip\bigskip

We acknowledge partial support from U.S. Israel BSF (G.E. and A.S.) 
and from the U.S. DOE contract numbers DE-AC02-98CH10886(BNL), 
DE-FG03-94ER40837(UCR). G.E. thanks the Israel Science Foundation and the Fund 
for the Promotion of Research at the Technion for partial support 
and members of the HEP group in UCR for their hospitality. 
\pagebreak


\begin{thebibliography}{99}
\bibitem{reffoot1} We neglect another possible lepton number 
violating term of the 
form $\mu L H_u$ which can induce sneutrino--Higgs-boson mixing   
and, in general, cannot be rotated away. We note that this term 
is of little relevance for our reactions since a possible 
sneutrino--Higgs-boson mixing will 
cause insignificant changes in our cross sections 
due to the smallness of the Higgs couplings to down-quarks 
and leptons.
\bibitem{foot1} Because of the stringent upper limits on the relevant 
 $\rp$ Yukawa couplings the $\hat t$-channel sneutrino exchanges yield a 
negligible contribution to the cross section compared to the 
resonance $\hat s$-channel sneutrino diagrams. The former are therefore 
being neglected throughout the paper.
\bibitem{foot2} We neglect  
the effects of electron-sneutrino exchanges in section 3 since 
the upper limits on the couplings $d_j \sneupm^e d_k$ are much smaller than 
$d_j \sneupm^\mu d_k$ \cite{rpreview}. 
Besides, as mentioned above, the rough limit 
$\Delta m_{\sneupm^e} / m_{\nu_e} \lsim {\rm few} 
\times 10^3$ obtained in \cite{mixyuval}, implies that 
electron-sneutrino mixing is expected 
to be much smaller then muon-sneutrino mixing.
\bibitem{footbound} Note that, in some instances, the bound 
$\Delta m_{\sneui} / m_{\nu_i} \lsim {\rm few} 
\times 10^3$ found in \cite{mixyuval} is relaxed by 
about an order of magnitude if one uses the laboratory bound 
for $m_{\nu_\mu}$. Given the state of the theoretical constructs 
we do not think that this is unreasonable and that 
evasion of this bound can be ruled out.
\bibitem{foottables} Note that some limits on the various 
$\lambda$ and $\lambda^{\prime}$ couplings scale with the masses
of sleptons and squarks rather than the sneutrinos masses, since 
they were derived from processes involving sleptons and squarks only.
For simplicity we assume here that these couplings
 possess the same scaling behavior also with the sneutrinos masses.
\bibitem{footsystematics} We caution that the contour shown in Figure 2 for 
$M_{\mu \mu}^-=50$ GeV is unlikely to be accessible to experiment 
as for this case $R^{\rm sneu} \lsim 10^{-2}$.
Such a small deviation from the SM will be very difficult 
to verify experimentally. The experimental problem should be 
significantly alleviated for $M_{\mu \mu}^-=100$ GeV 
and $M_{\mu \mu}^-=150$ GeV 
for which $R^{\rm sneu} > 0.1$ and $R^{\rm sneu} > 1$, respectively.
\bibitem{foot3} $T_N$ is 
defined as usual by replacing ${\rm time} \to - {\rm time}$ 
without changing initial and final states.
\bibitem{foottaupol} We want to emphasize that so far the 
$\tau$-spins have not been measured at the Tevatron and that  
the results of section 3 in this paper cannot be 
used until such a measurement is accomplished.
We are making an assumption that such a measurement with an 
efficiency of $\sim 10\%$ will be attainable in the future.
\end{thebibliography}
\end{document}